\documentclass[a4paper,11pt]{article}
\pdfoutput=1 % if your are submitting a pdflatex (i.e. if you have
\usepackage{jheppub} % for details on the use of the package, please
\usepackage[T1]{fontenc} % if needed

\usepackage{pdfsync}
\usepackage{mathptmx}
\usepackage[utf8]{inputenc}
\usepackage[T1]{fontenc}
\usepackage{slashed}
\usepackage{times}
\usepackage{amssymb,amsfonts,amsmath,amsthm}
\usepackage{dsfont,bbm}
\usepackage{dcolumn}
\usepackage{epsf}
\usepackage{graphicx}
\usepackage[caption=false]{subfig}
\usepackage{dsfont}
\usepackage{simplewick}
\usepackage{bm}
\usepackage{eucal}
\usepackage{slashed}
\usepackage[active]{srcltx}
\usepackage[usenames]{color}
\newcommand{\up}{{\uparrow}}
\newcommand{\down}{{\downarrow}}
\title{\boldmath TMD-like functions through the twisted quark states
}

\author[a, b, c]{I.~V.~Anikin,}
\author[b, c]{Xurong~Chen}
\affiliation[a]{Bogoliubov Laboratory of Theoretical Physics, JINR, Dubna}
\affiliation[b]{Institute of Modern Physics, Chinese Academy of Sciences, Lanzhou, P.R. China}
\affiliation[c]{Southern Center for Nuclear Science Theory,
Institute of Modern Physics, Chinese Academy of Sciences, Huizhou, P.R. China}

\emailAdd{anikin@theor.jinr.ru}

\abstract{
We investigate a new class of transverse momentum dependent functions (TMDs), 
as known as align-spin (AS) functions.
In the paper, we propose the most suitable proof of the AS-function existence together 
with the demonstration of the preponderances if   
the framework of twisted quark states has been employed.
The twisted state corresponds to the elementary particle (quark) which possesses 
the nontrivial intrinsic orbital angular momentum owing to
the swirling trajectory of motion.
In its turn, it leads to the cylindric system applied for the consideration.
In this connection, we reveal that the twisted (vortex) quark states serve as effective tools for the study of TMDs,
thereby facilitating a comprehensive analysis of AS-functions.
In contrast to the previous studies, where the existence of new TMDs is related to the the corresponding interactions 
encoded in the correlators,
we now focus on the leading order of interactions
providing a simplified and robust method.
This has been ensured by the twisted quark in the corresponding correlator because 
the essential transverse momentum dependence of quarks generated by interactions in the correlators 
can be alternatively described by the twisted states. 
Using a cylindrical formulation for twisted states, we can combine
the properties of plane-wave particles with a description stemmed from spherical harmonics,
resulting in well-defined propagation directions accompanied by essential OAM projections.
In particular, this innovative framework opens a new window for the direct investigations of AS-functions,
generating the unique angular $\phi$-dependence of differential cross sections.
It also points towards promising applications in experimental particle physics.
}

\begin{document}
\maketitle
\flushbottom
%%%%%%%%%%%%%%%%%%%%%%%%%%%%%%%%%%%%%%%%%%

%%%%%%%%%%%%%%%%
\section{Introduction}
\label{Intro}
%%%%%%%%%%%%%%%%%%%%%%%%%%%%%%%%%%%%

It is now well-known that both inclusive and exclusive hard reactions are important ways for extracting crucial insights
on the internal composite structure of hadrons.
Hard processes characterized by large momentum transfer are often regarded as a distinct class of interactions,
where the asymptotic regime permits a partial estimation and computation of the associated amplitudes,
as seen in Drell-Yan-like and Compton-like processes.
Numerous approaches have been
developed to address these phenomena, all of which fundamentally rely
on the factorization theorem
\cite{Boglione:1999pz, Bacchetta:2004jz, Goeke:2005hb, Collins:2005rq, Anselmino:2008sga, Bastami:2018xqd,
Boer:1997nt, Anikin:2022ocg}.
This theorem serves as a cornerstone for connecting the high-energy behaviour of scattering amplitudes
with the non-perturbative structure of the hadrons involved,
thereby enabling a deeper understanding of their internal dynamics.

\subsection{A short remark on factorization}

In the paper, the main object of application of the twisted particle conception is the transverse 
momentum dependent parton distributions. For further clarifications,
it is worth to remind the mathematical basis of any factorization 
\footnote{Every type of the parton distributions, strictly speaking, does not exist without factorization.}.
It  has been formed due to
the factorization theorem that has to be proven within the Standard Model (or QCD).
We can schematically demonstrate the result of factorization as
\begin{eqnarray}
\label{Math-conv-1}
{\cal A}/{\cal H} = \Big[E(x_1, ....) \otimes \Phi(x_1, ....)\Big]
\oplus
\Big[\Phi(y_1, ....) \otimes E(x_1, y_1; ....) \otimes \Phi(x_1, ....)\Big],
\end{eqnarray}
where ${\cal A}/{\cal H}$ is either the amplitude or the hadron tensor which are associated with
processes,
 $E(....)$ and $\Phi(....)$
denote the hard (perturbative) and soft (non-perturbative) parts, respectively.

We emphasize that the representation of ${\cal A}/{\cal H}$ in the form of
the mathematical convolution, see (\ref{Math-conv-1}), is entirely a consequence
of the corresponding asymptotical estimation. We may also say that the asymptotical estimation of the
given amplitudes/hadron tensors (instead of the direct calculations of amplitudes) leads to
the factorized mathematical convolution of the {\it physically independent} functions $E(x_1, ....)$
and $\Phi(x_1, ....)$.
By definition, the mathematical convolutions of any objects are given by
the {\it dimensionless} integrations over the corresponding manifolds.
Moreover, while the hard part structure, $E(x_1, ....)$, is fixed by the perturbative theory,
the soft part structure, $\Phi(x_1, ....)$, is determined by non-perturbative theory.
From the point of view of Standard Model,
$\Phi(x_1, ....)$ is not calculable object at all
\footnote{
We do not consider the effective models which are defined by the
phenomenological Lagrangians of quark-hadron interactions.}.
Instead, the properties of  $\Phi(x_1, ....)$ and its different representations which stem
from the fundamental symmetries and Lorentz covariance are currently being intensively studied.
The different functions, which are parametrizing the soft part function $\Phi(x_1, ....)$, can be treated as
the probability amplitudes (or their extensions) describing the different distributions of partons inside hadrons.
At the same time, the parton distributions can be connected with the experimental data.
Indeed, the experimental data analysis include the fitting parameters are tied with
the correlators of local Gegenbauer operators and can be traced from the parton distributions.

In the simplest case, we deal with the collinear, $k_\perp=0$ ($k_\perp$ is a primordial
transverse momentum), and non-interacting partons, $\mathbb{S}=1$
($\mathbb{S}$-matrix is trivial) in the corresponding correlators,
the parton distributions have the well-defined probability interpretations in terms of the corresponding probability amplitudes.
In this case, the parton distributions describe the ``static'' probability without any evolutions.

The evolution of parton distributions starts to be available
if one deals, at least, with the small $k_\perp$-dependence in the correlators
($|k_\perp|= {\cal O}(k^2_\perp)$)
and involves the interactions in the consideration ($\mathbb{S}\not=1$).
In this case, we still have the probability interpretations
of parton distributions.

However, the case of substantial $k_\perp$-dependence of the parton distributions (or paramtrizing functions)
destroys, as a rule, the interpretation of
$k_\perp$-dependent parton distributions as a probability amplitude.
But the probability interpretation can be restored only after the corresponding $k_\perp$-integration
(with the corresponding weight functions), {\it i.e.}  through the function moments.

In the present paper, we dwell on the
most attractive case of  $k_\perp\not=0$ (the substantial transverse momentum dependence) and
$\mathbb{S}\not=1$ (the presence of interaction in correlators). This case opens a new possibility
for the existance of new transverse momentum dependent distributions (TMDs).
In a series of papers \cite{Anikin:2021zxl, Anikin:2022omf,Anikin:2023krx,Anikin:2022eyf},
a new kind of TMDs has been introduced and described in detail. However, 
as the practice shows, the previous presentations 
suffer probably from the high level of detalization which may create artificial obstacles for understanding.
In this connection, it would be useful to give an analogous presentation in the most adopted way, shortly and clearly. 
The first part of the paper, see Sec.~\ref{Intro-AS}, devotes to this goal.

\subsection{A role of twisted particles in TMDs}

Recently, the twisted (vortex) states have been employed to elucidate various phenomena across different theoretical frameworks,
as exemplified in references \cite{Jentschura:2011ih, Ivanov:2022jzh, Bliokh:2017uvr, Serbo:2015kia}.
For example, the Mott scattering of high energetic twisted electrons on atoms has been studied 
within the Dirac relativistic theory discovering the fact that
the angular distribution of the outgoing electrons depends not only on the kinematic parameters of the twisted wave, but also on the total angular momentum.
A distinguishing feature of twisted states is their ability to effectively merge the characteristics
of traditional plane-wave particles with representations that incorporate spherical harmonics.
As demonstrated, the twisted states of particles (photons, electrons, and so) 
are non–plane wave solutions of the wave equation with helicoidal wave fronts. 
They possess, so to say, an intrinsic orbital angular momentum regarding 
the average propagation direction, see \cite{Ivanov:2022jzh} for the comprehensive study.
Consequently, twisted particles exhibit a well-defined propagation direction, accompanied by
essential orbital angular momentum (OAM) projections aligned along this propagation axis.
Unlike the conventional approaches, the treatment of OAM—considered within the context of total angular momentum
(TAM)—is enhanced by employing a conical momentum distribution, which incorporates angular phases
and is modulated by Bessel functions of the radial variable \cite{Jentschura:2011ih}.

Due to the above-mentioned properties, it turns out that 
the twisted particle conception can be an excellent instrument to explore 
the new (together, actually, with standard) TMDs.  
Indeed, as shown in the paper, the cumbersome reasonings, based on 
the inclusion of the corresponding interaction in correlators, to prove the existence of new TMDs 
can be  readily replaced by the rather trivial argumentation where there are no interactions provided
one of the quarks in the correlator has to be described as the twisted state.  
So, we present the additional evidence supporting a necessity of new TMDs through the utilization
of twisted (vortex) quark states in the corresponding correlators.

The presented framework not only enriches our understanding of the underlying dynamics but also underscores
the robustness of twisted quark states in revealing the presence of new TMDs.
The formalism of twisted quark states provides
a robust framework for the exploration of a new class of TMDs,
specifically align-spin (AS) functions.
We stress that the employment of twisted quark states not only enhances
the analytical capacity for examining these new TMDs
but also offers a methodology that can be adopted to study standard TMDs.
The second part of our paper, see Sec.~\ref{Tw-q}, involves the details concerning this topics.

\section{Introductory items of AS-functions}
\label{Intro-AS}

In this section, we provide the most needed information to demonstrate
the existance of new TMDs as known as the aligned-spin (AS) functions
(see \cite{Anikin:2021zxl, Anikin:2022omf,Anikin:2023krx,Anikin:2022eyf} for details).

\subsection{On the interactions in correlators}

To begin with, we remind that the hadron matrix elements (correlator) of the quark-gluon (non-local) operators
 contain always interactions via the presence of $\mathbb{S}$-matrix.
Indeed, any amplitudes and/or hadron tensors are determined by the hadron matrix elements of
${\rm T}$-product of the corresponding combinations of currents within the {\it interaction representation}.
For example, the forward Compton scattering (CS), $\gamma^*(q) + {\rm h}(P) \to
\gamma^{\, \prime}(q) + {\rm h}^{\, \prime}(P)$,
amplitude is given by
\footnote{Here and in what follows the brackets in the integral measure imply the
corresponding normalization constants which are irrelevant for our study.}
\begin{eqnarray}
\label{Amp-2}
{\cal A}_{\mu\nu} &=&
\langle P| a^-_\nu(q) \, \mathbb{S}[\bar\psi, \psi, A] \, a^+_\mu(q) | P\rangle
\nonumber\\
&=&
 \int (d^4 z) e^{- i q z} \langle P| \, {\rm T}
\Big\{ [\bar\psi(0)\gamma_\nu \psi(0)] \,
[\bar\psi(z)\gamma_\mu \psi(z)]\,
\mathbb{S}[\bar\psi, \psi, A]  \Big\}
| P\rangle,
%\nonumber
\end{eqnarray}
where the commutations of photon operators of annihilation and creation, $a^\pm_\mu$,
with the $\mathbb{S}$-matrix have been taken into account.  In (\ref{Amp-2}),
the hadron state
$ | P \rangle$ can be formally written  as $ \mathbf{a}^+_h(\psi, \bar\psi | A) \, | 0 \rangle$
with the hadron operator  $\mathbf{a}^+_h(\psi, \bar\psi | A)$ that remains to be undefined in pQCD.
Therefore, anticommutator $\big[ \, \psi(0), \, \mathbf{a}^+_h(\psi, \bar\psi | A)\, \big]_+$ is still unknown and
the direct calculation of amplitude is not available.

Focusing on the connected diagrams only and using
Wick's theorem, we derive the following contribution to the ``hand-bag''-type of  CS-diagrams
\footnote{$\delta^{(4)} (\text{\it momentum conserv.})$, as a common prefactor, is not shown.}:
\begin{eqnarray}
\label{Amp-3}
&&{\cal A}_{\mu\nu}^{\text{hand-bag}} =
\int (d^4 k) \,\text{tr} \big[ E_{\mu\nu}(k, q) \,\Gamma \big]  \, \Phi^{[\Gamma]}(k),
%\nonumber
\end{eqnarray}
where $\Gamma$ denotes the $\gamma$-matrix from the basis,
\begin{eqnarray}
%\label{E}
\label{phifun}
E_{\mu\nu}(k, q) = \gamma_\mu S(k+q) \gamma_\nu + (\text{\it cross.}),
\,\,
\Phi^{[\Gamma]}(k)= \int (d^4 z) \, e^{ikz}
\langle P| \,{\rm T} \, \bar\psi(0) \, \Gamma\,  \psi(z) \mathbb{S}[\bar\psi, \psi, A] | P \rangle.
%\nonumber
\end{eqnarray}
Instead of direct calculations,  ${\cal A}_{\mu\nu}$ can be estimated with the help of the suitable asymptotical regime (with a large virtuality of the initial photon,
$q^2=-Q^2 \to\infty$). The estimation of amplitudes within the asymptotical regime is a object of factorization procedure.
In order to compactify the representation of $\Phi(k)$, we can go over to the {\it Heisenberg representation} of correlators (see \cite{Anikin:2013gua} for the important details), {\it i.e.}
\begin{eqnarray}
\label{phifun-H}
\Phi^{[\Gamma]}(k) &=& \int (d^4 z) \, e^{ikz}
\langle P| \bar\psi(0) \, \Gamma\, \psi(z) | P \rangle^H
\nonumber\\
&\stackrel{{\cal F}}{=}& \langle \bar\psi(z) \, \Gamma \,\psi (0) \rangle,
%\nonumber
\end{eqnarray}
where the corresponding contour gauge has been used and the Wilson lines have been omitted
\cite{Anikin:2024vkl}.

Notice that, in what follows, according the traditions, if the symbol of ${\rm T}$-product is not shown,
the correlators will be always treated
within the Heisenberg representation even in the case where the symbol $H$ has been omitted.
We stress one feature of the Heisenberg representation: $\mathbb{S}$-matrix has been hidden in notations
but it does not mean the interactions are absent \cite{Anikin:2013gua}.

In the similar manner, we can consider the hadron tensor of any one-photon processes like
Drell-Yan (DY) process.
We have
\begin{eqnarray}
\label{Fact-1}
W=\int (d^4 k_1) (d^4 k_2) E(k_1, k_2, q;\,  \Gamma_1,  \Gamma_2) \,
\Phi_1^{ [\Gamma_1]}(k_1) \, \bar\Phi_2^{[ \Gamma_2]}(k_2),
%\nonumber
\end{eqnarray}
where
\begin{eqnarray}
\label{E-Phi}
&&E(k_1, k_2, q;\,  \Gamma_1,  \Gamma_2 ) = \delta^{(4)}(k_1+k_2-q) \, {\cal E}(k_1, k_2, q;\,  \Gamma_1,  \Gamma_2)
\nonumber\\
&&
\Phi_1^{ [\Gamma_1]}(k_1)\stackrel{{\cal F}_1}{=} \langle \bar\psi(z_1) \Gamma_1 \psi (0) \rangle,
\quad
\bar\Phi_2^{ [\Gamma_2]}(k_2)\stackrel{{\cal F}_2}{=} \langle \bar\psi(0) \Gamma_2 \psi (z_2) \rangle
%\nonumber
\end{eqnarray}
and $\stackrel{{\cal F}_i}{=}$ denotes again the corresponding Fourier transforms.

Notice that the presence of interactions in the correlators ensures the possibility of the evolutions
owing to the explicit loop integrations (in contrast to the implicit loop integrations), see below.

\subsection{The factorized amplitude/hadron tensor}
\label{FT:subsec}

In the preceding subsection, the non-factorized objects have been discussed, see (\ref{Amp-3}).
We currently focus on the results of factrorization procedure applied for the considered amplitudes/hadron tensors.

Since the factorization procedure, which we follow to, has been comprehensively described and discussed in many papers, see for example \cite{Anikin:2021zxl, Anikin:2022omf,Anikin:2023krx,Anikin:2022eyf}.
We give only the final factorized expressions for the amplitude/hadron tensor.
For the CS-like amplitude, we write
the following
\begin{eqnarray}
\label{Fact-2}
A^{(0)}=\int (d x) \, E(xP^+ ; q; \Gamma)
\Big\{
\int (d^4 k) \delta(x - k^{+}/P^{+})
\Phi^{ [\Gamma]}(k)
\Big\}
\end{eqnarray}
 if  $k^\perp_i$-terms are neglected in the expansion of $E(k, q)$;
 and
 \begin{eqnarray}
\label{Fact-3}
A^{(k_\perp)}=\int (d x) \sum_{i}E^{(i)}(xP^+ ; q ; \Gamma)
\Big\{
\int (d^4 k) \delta(x - k^{+}/P^{+}) \,
\prod_{i^\prime=1}^{i}
k_{i^\prime}^\perp\,
\Phi^{ [\Gamma]}(k)
\Big\}
\end{eqnarray}
if $k_\perp$-terms are essential in the expansion \cite{Anikin:2020ipg}.
For the DY-like hadron tensor, we obtain that
\begin{eqnarray}
\label{W-Fact-2}
&&W^{(0)}=\int (d x_1) (d x_2) E(x_1P^+_1, x_2P^-_2; q; \Gamma_1, \Gamma_2)
\\
&&
\times
\Big\{
\int (d^4 k_1) \delta(x_1 - k_1^{+}/P_1^{+})
\Phi_1^{ [\Gamma_1]}(k_1)
\Big\}
\Big\{
\int (d^4 k_2)  \delta(x_2 - k_2^{-}/P_2^{-})
\bar\Phi_2^{ [\Gamma_2]}(k_2)
\Big\}
\nonumber
\end{eqnarray}
for the unessential (integrated out in the soft functions) $k_\perp$-dependence;
 and
 \begin{eqnarray}
\label{W-Fact-3}
&&W^{(k_\perp)}=\int (d x_1) (d x_2) \sum_{i,j}E^{(i,j)}(x_1P^+_1, x_2P^-_2; q; \Gamma_1, \Gamma_2)
\\
&&
\times
\Big\{
\int (d^4 k_1) \delta(x_1 - k_1^{+}/P_1^{+}) \,
\prod_{i^\prime=1}^{i}
k_{1\, i^\prime}^\perp\,
\Phi_1^{ [\Gamma_1]}(k_1)
\Big\}
\Big\{
\int (d^4 k_2)  \delta(x_2 - k_2^{-}/P_2^{-})\,
\prod_{j^\prime=1}^{j}
k_{2\, j^\prime}^\perp\,
\bar\Phi_2^{ [\Gamma_2]}(k_2)
\Big\}
\nonumber
\end{eqnarray}
for the essential $k_\perp$-dependence in the soft functions.

\subsection{Introducing AS-functions}

As well-known, every kinds of parton distribution functions are nothing but the Lorentz parametrization
of relevant correlators obtained from the factorization procedure which has been applied to the
amplitudes/hadron tensors.
In its turn, the Lorentz parametrizations are being implemented in terms of the external and internal
Lorentz vectors (or tensors). By definition, the external tensors are generated by the hadron characteristics
such as momentum, spin and so on, while the internal tensors are accumulating the quark-gluon
(or parton) characteristics.
As demonstrated in \cite{Anikin:2022omf}, the set of internal parton Lorentz vectors has been formed by the
corresponding spinor lines.

Indeed, let us begin with the simplest case of
$k_\perp=0$ and $\mathbb{S}=1$.
In this case, for the CS-amplitude, see (\ref{Amp-3}),
we have the following
\begin{eqnarray}
\label{Amp-2-2}
{\cal A}_{\mu\nu}\Big|_{\mathbb{S}=1} =
\int (d^4 z_1\, d^4 z_2) \, e^{- i q (z_1 - z_2)} \langle P| \,
 \textbf{:} \bar\psi(z_1) \, E_{\mu\nu}(z_1-z_2) \,\psi(z_2)  \textbf{:} \,
| P\rangle.
%\nonumber
\end{eqnarray}
in the co-ordinate space, and,
focusing on $\gamma^+$-projection in the correlator,
\begin{eqnarray}
\label{Amp-3-2}
{\cal M}_{\mu\nu} =
\int (d^4 k) \,\text{tr} \big[ E_{\mu\nu}(k) \, \gamma^- \big]
\Phi^{[\gamma^+]}(k)
\end{eqnarray}
with
\begin{eqnarray}
\label{Phi-N-prod}
\Phi^{[\gamma^+]}(k)  \Big|_{\mathbb{S}=1}=
\int (d^4 z) \, e^{ikz}
\langle P| \,   \textbf{:}  \bar\psi(0) \gamma^+ \psi(z)  \textbf{:} \,| P \rangle
%\nonumber
\end{eqnarray}
within the momentum representation.
On the other hand,
having used the Fourier transforms for the operators in the correlator, we can readily derive that
\footnote{For the sake of shortness, the contribution of the anti-quark combination is not displayed.}
\begin{eqnarray}
\label{Amp-3-3}
\langle P| \,   \textbf{:}  \bar\psi(0) \gamma^+ \psi(z)  \textbf{:} \,| P \rangle
= \int (d^4 k_1 d^4 k_2) e^{ -i k_1 z}
L^{[\gamma^+]}(k_2, k_1) \,\,
\langle P | b^+(k_2) b^-(k_1) | P \rangle,
%\nonumber
\end{eqnarray}
where $L^{[\gamma^+]}(k_2, k_1)$ gives the internal Lorentz tensor needed for  the parametrization
and it is related to the spinor line  $\big[ \bar u(k_2) \gamma^+ u(k_1) \big]$.
In (\ref{Amp-3-3}),  $\langle P | b^+(k_2) b^-(k_1) | P \rangle$ defines
the quark-hadron ${\cal M}$-amplitude, $\delta^{(4)}(k_1-k_2) {\cal M}(k_2,k_1 ;\, P)$.

We stress that,  after factorization at the leading order and in the collinear limit,
the correlator can be finally treated through the mathematical probability to find a parton inside the
given hadron, {\it i.e.}
\begin{eqnarray}
\Phi^{[\gamma^+]}(x) \Big|_{\mathbb{S}=1} = \int (d^4 k) \delta(x-k^+/P^+) \Phi^{[\gamma^+]}(k)
\stackrel{{\cal F}}{=}
\langle P| \,   \textbf{:}  \bar\psi(0) \gamma^+ \psi(0^+,z^-, \vec{\bf 0}_\perp)  \textbf{:} \,| P \rangle.
%\nonumber
\end{eqnarray}
where $k=(k^+, k^-, \vec{\bf k}_\perp)$.

The next stage is to include the interaction in correlators.
As explained in  \cite{Anikin:2022omf}, the essential $k_\perp$-dependence of any
parton distributions ($k_\perp$-unintegrated functions), as a rule, stems from the interaction encoded in the correlator.
So, we have
(here, the notation $\tilde z=(z^-, \vec{\bf z}_\perp)$ has been used)
\begin{eqnarray}
\label{Phi-T-prod}
\hspace{-0.2cm}\Phi^{[\gamma^+]}(x, k_\perp) = \int (dk^+ dk^-) \delta\big(x-k^+/P^+\big) \Phi^{[\gamma^+]}(k)
\stackrel{{\cal F}}{=}
\langle P| \,{\rm T} \, \bar\psi(0) \gamma^+ \psi(0^+, \tilde z)
\mathbb{S}[\bar\psi, \psi, A] | P \rangle.
%\nonumber
\end{eqnarray}
As above-mentioned,  the correlator in the form of (\ref{Phi-T-prod}) appears naturally in the
amplitude/hadron tensor at the given order, see (\ref{Amp-3}). At the same time, the probability
interpretation of factorized correlators is now not available owing to the substantial $k_\perp$-dependence
unless the integrations over $k_\perp$ have been performed, see (\ref{Fact-3}) and (\ref{W-Fact-3}).

The correlator that defines the function $\Phi^{[\gamma^+]}(x, k_\perp)$ possesses the external Lorentz index
$\mu=+$ (within the light cone basis) together with the external Lorentz tensor $P$
(the hadron momentum). Generally speaking, the set of external Lorentz tensors is related to the hadron characteristics.
For example, the hadron spin $S$ can be involved in the external set too.

Moreover, in $\Phi^{[\gamma^+]}(x, k_\perp)$,
the loop integrations, which are appeared at the given order of interactions using Wick's theorem,
give both the explicit (evolution)
and the implicit (structure) integrations.
By definition, the explicit loop integration contains the integration with the closed spinor loop (the product
of fermion propagators). In contrast to that, the implicit integration does not form the closed spinor circle.
As a result, only the explicit loop integration can generate the evolution of the corresponding operator,
while the implicit integration is responsible for the spinor lines which correspond to the different Lorentz tensors.

Concerning the set of internal Lorentz tensors,
having fixed the order of $g$ (the coupling constant) in the $\mathbb{S}$-matrix
expansion, we are able to form the needed set of internal Lorentz tensors which parametrize the
correlator together with the set of external Lorentz tensors.
We remind that, as a rule, the internal Lorentz set  have been generated by the corresponding spinor lines.
Indeed, at the order of $g^2$, we have the following (see, for example,  \cite{Anikin:2022omf})
\begin{eqnarray}
\label{Me-2}
&&
\langle P,S | T \bar\psi(0) \gamma^+ \psi(z) \, \mathbb{S}_{QCD}^{(2)}[\psi,\bar\psi, A] \,| P,S \rangle
\Big|_{\text{implicit loop integr.}} \sim
\nonumber\\
&&
\big[ \bar u (k) \gamma^+ \hat k \gamma^\perp_\alpha u(k-\ell)\big]
\big[ \bar u (\tilde k) \gamma^\perp_\alpha u(\tilde k + \ell)\big],
%\nonumber
\end{eqnarray}
where $\sim$ implies ``involves''.
Assuming the kinematic regime: $| \ell | \ll \{|k|, |\tilde k| \}$ and $|\tilde k| \sim |k|$,
after a simple spinor algebra based on the Fierz transformations (Fi. Tr),
we can derive that (see \cite{Anikin:2021zxl, Anikin:2022ocg, Anikin:2022eyf} for all details)
\begin{eqnarray}
\label{Me-2w}
&&
\langle P,S | T \bar\psi(0) \gamma^+ \psi(z) \, \mathbb{S}_{QCD}^{(2)}[\psi,\bar\psi, A] \,| P,S \rangle
\Big|_{\text{implicit loop integr.}}
\nonumber\\
&&
\stackrel{{\rm Fi. \, Tr.}}{
\Longrightarrow}
\big[ \bar u^{(\up_x)}(k) \gamma^+ \gamma^\perp \gamma_5 u^{(\up_x)}(k)\big]
\big[ \bar u^{(\up_x)}(k) u^{(\up_x)}(k)\big] \sim s_\perp.
\end{eqnarray}

At the order of $g^4$, {\it i.e.} $\mathbb{S}^{(4)}_{QCD}$-term in expansion, the appearance of the relevant spinor line 
associated with the quark spin can be observed even within the simple spinor algebra.
In this case, there are no a necessaty for the Fierz transforms and the special kinematical regimes, 
see \cite{Anikin:2022omf, Anikin:2023krx}.

As a result, the quark spin (axial)vector $s_\perp$ appears as the consequence of inner
interactions encoded in the correlator and it has to be included in the set of internal Lorentz tensors
to parametrize the correlator,
{\it i.e.}
\begin{eqnarray}
\label{Me-2w-2}
&&
\langle P, S|  T \bar\psi(0)\,\gamma^+\, \psi(z)\,\,
\mathbb{S}^{(2)}_{QCD}[\bar\psi, \psi, A] |P, S \rangle
\Big|_{\text{implicit loop integr.}}
\stackrel{{\cal F}}{=}
i \epsilon^{+ - P_\perp s_\perp} \tilde f_1^{(1)} + ....
\end{eqnarray}

Notice that if $|k_\perp|= {\cal O}(k^2_\perp)$, hence it influences on the explicit and implicit integrations.
 Namely, the implicit integration becomes trivial and the explicit integration gives
 the non-trivial evolution kernel.
 So, in this case, the implicit loop integration cannot produce the new type of parton distributions with
the essential $k_\perp$-dependence.
In other words, we can conclude that, in the case of small transversities, the Lorentz
parametrization procedure and the explicit (evolution) loop integration
``commutes'' with one another.

However, if $|k_\perp|\not= {\cal O}(k^2_\perp)$,
the explicit (evolution) loop integration and the Lorentz parametrization are not
``commutative''. That is, we have first to implement the parametrization and, then, to study the corresponding evolutions.
Besides, in the $k_\perp$-dependent case, the implicit (structure) integration gives more possibilities for the parametrization
which ultimately lead to the presence of new functions.

\section{The twisted (vortex) quark state in the correlators}
\label{Tw-q}

In the previous section, we outline the main evidences for that the new TMDs exists and extends
the standard set of transverse momentum dependent parton distributions.
We have demonstrated that, at the order of $g^2$, using the Fierz transformations
the additional spinor lines appear and they generate
the needed quark transverse spin axial-vector for the Lorentz parametrization.
However, there is the simplest way to observe the new TMDs thanks to the use of twisted particles.

Before going further, we make two observations.
First, one can see that
if one fixes the spinor polarization along $x$-axis,
the operator $\bar\psi(0)\,\gamma^+\, \psi(z)$ defining
$\Phi^{[\gamma^+]}(x, k_\perp)$ can be written as
\begin{eqnarray}
\label{OP-s}
\bar\psi^{(\up_x)} \,\gamma^+\, \psi^{(\up_x)} =
\bar\psi^{(\up_x)} \,\gamma^+\gamma_1 \gamma_5\, \psi^{(\up_x)},
\end{eqnarray}
where $\psi^{(\up\down_i)}=1/2 (1 \pm \gamma_i \gamma_5) \psi$ with $i=(1,2)\equiv (x, y)$.

Second, from (\ref{Me-2}), based on the simple algebra, one can conclude that the spinor line defined by
$\big[ \bar u (k) \gamma^+ \hat k \gamma^\perp_\alpha u(k)\big]$ can be reduced to the spinor line
such as $k_\alpha^\perp\big[ \bar u (k) \gamma^+ u(k)\big]$. On the other hand, as well-known
the presence of $k_\perp$ in the spinor line or in the correlator signals on the non-trivial contributions of OAM
(see for example \cite{Anikin:2015ita}).
The excellent way to include OAM in the correlator is to use the conception of the twisted particles
\cite{Jentschura:2011ih, Ivanov:2022jzh, Bliokh:2017uvr, Serbo:2015kia}.
In this case, we have a possibility to study the new kind  of TMDs even at the leading order level, see below.

Having said that, we go over to consideration of the correlator where one of quarks
has been replaced on the twisted quark. We begin with the leading order of $\mathbb{S}$-expansion,
we have
\begin{eqnarray}
\label{Me-Tw-1}
\Phi_{{\rm TW}}^{[\gamma^+]}(x, k_\perp)
\stackrel{{\cal F}}{=}
\langle P, S|  \bar\psi(0)\,\gamma^+\, \Psi_{{\rm TW}}(z)\, |P, S \rangle^{l_z\not=0},
\end{eqnarray}
where $\Psi_{{\rm TW}}(z)$ denotes the twisted (vortex) quark state and $\mathbb{S}=1$, for the moment.

\subsection{The twisted quark states in a nutshell}

For the pedagogical reason, it is worth to recall the standard spherical $SL(2, C)$-spinor represented as
( $\stackrel{\text{\tiny C-G}}{\leadsto}$ means ``modulo Clebsh-Gordon's coefficient'')
\begin{eqnarray}
\label{Sph-sp-1}
\psi_{J\, M}
\stackrel{\text{\tiny C-G}}{\leadsto}
\big[ R_{k\, l}(r) \, Y_{l\, l_z} (\theta, \varphi) \big]
 \circledast \varphi_{p \theta \phi,\, \lambda},
\end{eqnarray}
where $\hat J=\hat L+ \hat S$, $\hat J_3\to  M = l_z + s_z$, $\lambda$ is an eigenvalue of spin operator, and
$\vec{p}=(p_x, p_y, p_z)\equiv (p, \theta, \phi)$ with
\begin{eqnarray}
\label{p-sher-s}
\big\{ p_x=p\,\cos\phi\, \sin\theta, \quad p_y=p\,\sin\phi\, \sin\theta, \quad
p_z=p\, \cos\theta \big\}.
\end{eqnarray}

In (\ref{Sph-sp-1}), as usual, the OAM-part, $ R_{k\, l}(r) \, Y_{l\, l_z} (\theta, \varphi)$,
and the SAM, $\varphi_{p \theta \phi,\, \lambda}\equiv \varphi_{p \theta \phi,\, \lambda} (\bar\theta, \bar\phi)$,
have been well-separated independently the spin-orbital interactions
\footnote{Generally speaking, the angular dependence, $(\theta, \phi)$, of the Lorentz boost differs
from the angular dependence, $(\bar\theta, \bar\phi)$, of the spin quantization axis.}.

The OAM-part defined by the radial and spherical functions are given by
\begin{eqnarray}
\label{Sph-Har}
R_{k\, l}(r)= \sqrt{\frac{2\pi\, k}{r}} \, J_{l+1/2}(kr), \quad
Y_{l\, l_z} (\theta, \phi)  = \Theta_{l\, l_z} (\theta) \, e^{i l_z \phi}.
\end{eqnarray}

It is important to stress that, in QFT where the quark is a point-like particle, 
the factorized (separated) form of (\ref{Sph-sp-1}) has been dictated by the fact that 
the OAM-operator commutes with SAM-operator, $[ \,\hat L,\, \hat S\, ]=0$, even if the 
spin-orbital interaction, which becomes rather sizeable in nuclei, is presented. 
Indeed, the realization of the fictive internal space rotation giving the representation of $\hat S$-operator does not touch 
the rotation in $\mathbb{R}^3$ related to $\hat L$-operator. 
In other words, the generator that determines $\hat S$-operator excludes the space coordinate dependence, otherwise it would be impossible 
to construct the Pauli-Lubansky vector $W_\alpha$ which is one of the Poincar\'e group characters.   
In this connection, the representation of (\ref{Sph-sp-1}) can be treated as the state function of two particles in the {\it c.m.s.},
one of them is a scalar particle and it has the non-zero OAM, while the other one is a spinor with the zero OAM.
This is a reason for the Clebsh-Gordon coefficients. 

Based on the above-mentioned argumentation, we are readily able to describe the quarks in the cylindric system.  
To demonstrate this, we first have to recall that, 
as shown in \cite{Jentschura:2011ih}, 
the scalar state in cylindric system can be described by the 
wave function $\Phi_{\varkappa m k_z}$ which satisfies the Klein-Fock-Gordon equation (with the zero mass)
and involves the corresponding Bessel function.
Let us now go over to the  twisted (swirling) quark  
\cite{Jentschura:2011ih, Ivanov:2022jzh, Bliokh:2017uvr, Serbo:2015kia}, {\it i.e.} the quark 
propagates along the $z$-axis and have well-defined values of the longitudinal linear momentum $k_z$. 
Moreover, we have  
the modulus of the transverse momentum $|k_\perp|= \varkappa $ and 
the half-integer projection of the total angular momentum. 
Therefore, using the factorized form of (\ref{Sph-sp-1}), it is enough to make the 
replacement of spherical functions on the cylindric function to describe the swirling quark:
\begin{eqnarray}  
\label{Cyl-fun-1}
\big[ R_{k\, l}(r) \, Y_{l\, l_z} (\theta, \phi) \big] 
\stackrel{{\rm repl.}}{\Longrightarrow} 
\sqrt{\frac{\varkappa}{2\pi}} J_{l_z}(\varkappa r) \, e^{i l_z \phi}\Big|_{\text{twisted}}
\end{eqnarray}
and, in its turn, in the momentum representation, we have
\begin{eqnarray}  
\label{Cyl-fun-2}
\sqrt{\frac{\varkappa}{2\pi}} J_{l_z}(\varkappa r) \, e^{i l_z \phi}
\stackrel{\text{$p$-sp.}}{\Longrightarrow} 
e^{i l_z \phi} \sqrt{\frac{2\pi}{\varkappa}} 
\delta\big( | \vec{\bf k}_\perp| - \varkappa \big).
\end{eqnarray}

We now concentrate on the spinor part of $\psi_{J\, M}$.
At the beginning, for the pedagogical reason, we dwell on the massive quarks.
Then, due to the features of factrorization, we go over to the massless case
where the left and right Weyl $SL(2, C)$-spinors are independent ones.
Since we follow to the Wigner helicity method, both cases are not drastically different.

\textbf{The massive quarks.}
As mentioned in (\ref{OP-s}), it is necessary to extract the fermion (quark) states
with the transverse polarization, $\psi^{(\up_x)}$. For this goal,
we assume the $x$-axis to be
played a role of the spin quantization axis. The general representation of
the spin quantization axis is given by
 $\vec{n}(\bar\theta, \bar\phi)$ as a function of angles.
That is, we fix the angular dependence of the original (at the rest frame)
$SL(2, C)$-spinor as  $\varphi_{000, \lambda}(\bar\theta=\pi/2, \bar\phi=0)$, {\it i.e.}
\begin{eqnarray}
\label{RF-sp-1}
\varphi_{000, +\frac{1}{2}}\Big(\bar\theta=\frac{\pi}{2}, \bar\phi=0\Big)=
    \begin{pmatrix}
    1\\
    1
    \end{pmatrix},
\quad
\varphi_{000, -\frac{1}{2}}\Big(\bar\theta=\frac{\pi}{2}, \bar\phi=0\Big)=
    \begin{pmatrix}
    -1\\
    1
    \end{pmatrix}.
\end{eqnarray}
Then, we implement the Lorentz boost along $z$-axis, $\vec{p}=(0, 0, p_z)\equiv (p, 0, 0)$.
In other words, the spin quantazation axis differs from the
moving direction (boost) of particle. That is, we have
\begin{eqnarray}
\label{Bo-sp-1}
\varphi_{p00, \pm\frac{1}{2}}\Big(\bar\theta=\frac{\pi}{2}, \bar\phi=0\Big)
\stackrel{N}{=}
    \begin{pmatrix}
    \pm (E+m + p_z)\\
    (E+m - p_z)
    \end{pmatrix},
\end{eqnarray}
where, for the sake of shortness,
the normalization factor given by $[2m (E+m)]^{-1/2}$ has been absorbed in the symbol $\stackrel{N}{=}$.
The state $\varphi_{p00, \lambda}(\pi/2, 0)$ of (\ref{Bo-sp-1}) can be re-expressed through the helicity states
$\varphi_{p00, \lambda}(0, 0)$ as
\begin{eqnarray}
\label{Bo-sp-2}
\varphi_{p00, \pm\frac{1}{2}}\Big(\bar\theta=\frac{\pi}{2}, \bar\phi=0\Big)
= \varphi_{p00, -\frac{1}{2}}\Big(\bar\theta=0, \bar\phi=0\Big)
\pm \varphi_{p00, +\frac{1}{2}}\Big(\bar\theta=0, \bar\phi=0\Big).
\end{eqnarray}
With the help of (\ref{Bo-sp-2}) , one can readily obtain that
\begin{eqnarray}
\label{Bo-sp-3}
&&
 \varphi_{p\theta \phi, \pm\frac{1}{2}}\Big(\bar\theta=\frac{\pi}{2}, \bar\phi=0\Big)\equiv
U\Big( {\cal R}(\phi, \theta, 0) \Big) \, \varphi_{p00, \pm\frac{1}{2}}\Big(\bar\theta=\frac{\pi}{2}, \bar\phi=0\Big)
=
\nonumber\\
&&
\varphi_{p\theta \phi, -\frac{1}{2}}\Big(\bar\theta=0, \bar\phi=0\Big)
\pm \varphi_{p\theta \phi, +\frac{1}{2}}\Big(\bar\theta=0, \bar\phi=0\Big),
\end{eqnarray}
where $U\big( {\cal R}(\phi, \theta, 0) \big)$ denotes the rotation operator which is determined on the spinor
representation.
Hence,  (\ref{Bo-sp-3}) shows how the fermion state with the transverse polarization can be expressed through the
corresponding helicity states,
\begin{eqnarray}
\label{Bo-sp-4}
&&
 \varphi_{p\theta \phi, \pm\frac{1}{2}}\Big(\bar\theta=\frac{\pi}{2}, \bar\phi=0\Big)
\leadsto \psi^{(\up_x)},
\end{eqnarray}
which helps to prepare the corresponding twisted quark state.

\textbf{The massless quarks.}  Since we deal with TMDs (as are any kinds of parton distributions)
which have been arisen from the factorization procedure,
see Subsec.~\ref{FT:subsec}, the quarks should be considered as massless objects.
In this connection, the above-presented discussion on the massive spinors can be readily
reduced to the massless quark case. The only difference is that there is no the rest system for the
massless particle. Hence, the original $SL(2, C)$-spinor that has been above defined as
$\varphi_{000, \lambda}(\bar\theta=\pi/2, \bar\phi=0)$  should be replaced on the other
$SL(2, C)$-spinor with the fixed momentum $\tilde p$, $\varphi_{\tilde p 00, \lambda}(\bar\theta=\pi/2, \bar\phi=0)$.
The latter $SL(2, C)$-spinor should be then modified under the Lorentz boost if there is a need to change momentum
along $z$-axis direction,
\begin{eqnarray}
\label{L-b-m0-1}
U\big( {\cal L}(p, 0, 0)\big)\varphi_{\tilde p 00, \lambda}\big(\bar\theta=\frac{\pi}{2}, \bar\phi=0 \big) =
\varphi_{p 00, \lambda}\big( \bar\theta=\frac{\pi}{2}, \bar\phi=0\big),
\end{eqnarray}
where $U\big( {\cal L}(p, 0, 0)\big) = \exp\{ \sigma_3 p\, \tilde\varphi/2\} $ denotes the Lorentz boost along $(p, 0, 0)$.
Afterwards, we can use the transformations of (\ref{Bo-sp-2}) and (\ref{Bo-sp-3}).
It is worth to mention that the massless quark case leads to the independent $(\frac{1}{2}, 0)$- and $(0, \frac{1}{2})$-spinors,
the left and right Weyl spinors respectively.
Indeed, in order to compensate the nullification of denominators in the normalization constants, see (\ref{Bo-sp-1}),
we have to suppose that
\begin{eqnarray}
\label{M0-spin-1}
\big[ E- \vec{\sigma} \vec{p}\big] \varphi^{(R)}=0, \quad  \big[ E+ \vec{\sigma} \vec{p} \big] \varphi^{(L)}=0
\end{eqnarray}
independently. In its turns, (\ref{M0-spin-1}) immediately leads to the helicity states
\footnote{We remind that the helicity operator for the massless case is the Lorentz invariant operator.}:
$\varphi^{(R)}\equiv \varphi_{p\theta \phi, +1/2}$ and $\varphi^{(L)}\equiv \varphi_{p \theta\phi, -1/2}$.

Given that, we derive the twisted quark state with the transverse polarization in the form of
(we adhere the Weyl representations for spinors)
\begin{eqnarray}
\label{Psi-Tw-1}
\Psi^{(\up_x)}_{{\rm TW}}(z)=
\int (d^2k_L) (d^2 \vec{\bf k}_\perp) a_{\varkappa\, l_z}( \vec{\bf k}_\perp)\, e^{-ikz}
u^{(\up_x)}(k) \, b_{(\up_x)}(k)
+ (\text{anti-quark term}),
\end{eqnarray}
where the weight function
\begin{eqnarray}
\label{A-tw-1}
a_{\varkappa\, l_z}( \vec{\bf k}_\perp)=(-i)^{j_z} e^{i l_z \phi} \sqrt{\frac{2\pi}{\varkappa}}
\delta\big( | \vec{\bf k}_\perp| - \varkappa \big)
\end{eqnarray}
has been inspired by the cylindric frame for the twisted particle, see (\ref{Cyl-fun-1}) and (\ref{Cyl-fun-2}).
Moreover, in (\ref{Psi-Tw-1}), having used (\ref{Bo-sp-3}) and (\ref{Bo-sp-4}), one can see that
\begin{eqnarray}
\label{H-st-1}
u^{(\up_x)}(k) \sim \sum_{\lambda^\prime=\pm 1/2}
e^{-i \phi \lambda^\prime} \, d_{+\frac{1}{2} \lambda^\prime} (\theta) w^{(\lambda^\prime)}
\quad \text{with}\quad
w^{(\pm \frac{1}{2})} \equiv \varphi_{000, \pm \frac{1}{2}}(0, 0),
\end{eqnarray}
where $d_{\lambda \lambda^\prime} (\theta)$ implies the Wigner $D$-function.

It is important to emphasize that in (\ref{A-tw-1}) one has the eigenfunction of $\hat L_z$-operator ($z$-projection of OAM)
which is $\Phi_{l_z}(\phi)=e^{i l_z \phi}$. In the momentum space, the weight function $a_{\varkappa\, l_z}( \vec{\bf k}_\perp)$
adopts the index $l_z\pm \lambda$ only after the expansion of $u(k)$-spinor over the helicity functions
(pure spin states) $w^{(\lambda)}$, see (\ref{H-st-1}). That is, we write the following
\begin{eqnarray}
\label{Psi-Tw-2}
\Psi^{(\up_x)}_{{\rm TW}}(z)&=&
\int (d^2k_L) (d^2 \vec{\bf k}_\perp) \, e^{-ikz}
 \sum_{\lambda^\prime=\pm 1/2}  a_{\varkappa\, l_z\mp\lambda^\prime}( \vec{\bf k}_\perp)
\, d_{+\frac{1}{2} \lambda^\prime} (\theta) w^{(\lambda^\prime)}  \, b_{(\up_x)}(k)
\nonumber\\
&+& (\text{anti-quark term}).
\end{eqnarray}
This expression shows directly the source of the imaginary part of $\Psi^{(\up_x)}_{{\rm TW}}(z)$
which is induced by the phase factor of
the weight function $a_{\varkappa\, l_z\mp\lambda^\prime}( \vec{\bf k}_\perp)$ related to the
twisted particle.

\subsection{The correlator with twisted quark state}

Let us return to the correlator (\ref{Me-Tw-1}), we are now inserting
the representation given by (\ref{Psi-Tw-1}) into the correlator and we finally obtain that
\begin{eqnarray}
\label{Me-Tw-2}
&&
\langle P, S|  \bar\psi(0)\,\gamma^+\, \Psi^{(\up_x)}_{{\rm TW}}(z)\, |P, S \rangle^{l_z\not=0}=
\int (d^4 p) \int (d^2k_L) (d^2 \vec{\bf k}_\perp) a_{\varkappa\, l_z}( \vec{\bf k}_\perp)\, e^{-ikz}
\nonumber\\
&&\times
\big[ \bar u^{(\lambda^\prime)} (p) \,\gamma^+ \, u^{(\up_x)}(k)\big]\,\,
\langle P, S|  b^+_{(\lambda^\prime)}(p)\, b_{(\up_x)}(k) |P, S \rangle
+ (\text{anti-quark term}).
\end{eqnarray}
Notice that the fixed transverse polarization for the twisted quark singles out also the transverse polarization
for the standard quark. On the other hand, the twisted quark state has a unique dependence on the azimuthal angle
$\phi$ which is related to the TAM, see (\ref{A-tw-1}) and (\ref{H-st-1}), and is appeared owing the phase factor.
In this sense, if we study the $\phi$-dependence of the considered correlator contribution
to the given differential cross section,
we are able to extract information on new kind of TMDs generated by (\ref{Me-2w-2}).
This is one of the principle conclusions of the presented paper.

As above-mentioned, we can associate the phase factor of the twisted quark with the imaginary part source related to
the considered correlator and, therefore, to
the parametrizing function. This is a minimal but enough source of the corresponding complexity, see also \cite{Anikin:2022eyf}.

To demonstrate the advantages of the twisted particle conception, let us consider the
unpolarized Drell-Yan (DY) process, {\it i.e.}
\begin{eqnarray}
\label{DY-proc-st}
N(P_1) + N(P_2) \to \gamma^*(q) + X(P_X)
\to\ell(l_1)+\bar\ell(l_2) + X(P_X),
\end{eqnarray}
with the initial unpolarized nucleons $N$. This is the lepton-production in nucleon-nucleon collisions.
The importance of the unpolarized DY differential cross section is due to
the fact that it has been involved in the denominators of any spin asymmetries.

Following \cite{Anikin:2021zxl}, we write down the unpolarized differential cross section as
\begin{eqnarray}
\label{xsec-unpl}
d\sigma^{unpol.} &\sim& \int (d^2 \vec{\bf q}_\perp) {\cal L}^{U}_{\mu\nu}
{\cal W}^{(0)}_{\mu\nu} =
\int (d x)  (d y) \delta(x P^{+}_1 - q^+)
\delta(yP^{-}_2 -q^-)
\nonumber\\
&\times&
(1+\cos^2\theta) f(y)  \int (d^2 \vec{\bf k}_1^\perp ) \epsilon^{P_2 - k_1^\perp s^\perp} \Im{m}
f^{({\rm TW})}_{(2)} (x; \,k^{\perp\, 2}_1),
\end{eqnarray}
where $ f(y)$ is the standard parton distribution function parametrizing one of correlators, $\Phi^{[\gamma^-]}(y)$,
in the DY-hadron tensor and
$\epsilon^{+ - k_1^\perp s^\perp} = \vec{\bf k}_1^\perp \wedge \vec{\bf s}^{\perp} \sim \sin (\phi_k - \phi_s)$
with $\phi_A$, for $A=(k, s)$, denoting the angles between $\vec{\bf A}_\perp$ and $O\hat x$-axis in the
Collins-Soper frame.
The function $f_{(2)} (x; \,k^{\perp\, 2}_1)$ refers to the new kind of TMDs and it parametrizes the correlator
in the form of (cf. (\ref{Me-2w-2}) where the given TMDs is not shown explicitely)
\begin{eqnarray}
\label{DY-funs}
\bar\Phi^{[\gamma^+]}_{{\rm TW}}(x, k^{\perp\,2}_1) =
i\epsilon^{+ - k_1^\perp s^\perp} f^{({\rm TW})}_{(2)} (x; \,k^{\perp\, 2}_1).
\end{eqnarray}
In contrast to the previous cases \cite{Anikin:2021zxl, Anikin:2022omf,Anikin:2023krx,Anikin:2022eyf},
in this paper
the imaginary part of $f_{(2)} (x; \,k^{\perp\, 2}_1)$  of (\ref{DY-funs}) is determined by the phase factor of the twisted quark
which is involved in the correlator.

To conclude we notice that the angle $\phi_s$ cannot explicitly be measured in the experiment.
However, the implementation of the covariant (invariant) integration of $f_{(2)} (x; \,k^{\perp\, 2}_1)$  gives the
kinematical constraints on this angle relating the quark spin angle to the corresponding hadron angle
\cite{Anikin:2021zxl, Anikin:2022omf,Anikin:2023krx,Anikin:2022eyf}.

\section{Conclusions}

In the first part of paper, we have proposed the most suitable proof of the existence of new TMDs
introduced in \cite{Anikin:2021zxl, Anikin:2022omf,Anikin:2023krx,Anikin:2022eyf}.
From the practical point of view, it becomes actual because 
the previous presentations are very much detailed and generated 
the artificial obstacles for understanding.

In the second part of paper, we have 
demonstrated that the framework of twisted quarks 
\cite{Jentschura:2011ih, Ivanov:2022jzh, Bliokh:2017uvr, Serbo:2015kia} 
serves as a highly effective approach for investigating a new class of transverse
momentum distributions (TMDs),
specifically the align-spin (AS) functions.
The proposed approach can be also adopted to study the standard TMDs.

As shown in the paper, with
 the help of the twisted quark, it is enough to be limited by the leading order of interaction
to observe AS-functions. This is a simplest and more realible way compared to the traditional methods
which are based on the $\mathbb{S}$-matrix expansion.
Indeed, the essential transverse momentum dependence can be traced not only from the corresponding
interaction but also from the non-trivial OAM contributions which are, in its turns,
induced by the transverse momentum dependence of correlators
\footnote{Roughly speaking, the needed covariant derivative is always a sum of
the transverse partial derivative (giving the non-trivial $k_\perp$-dependence in the momentum representation)
and the transverse gluon term.}.
Since, the twisted states, defined within the cylindric frame, reflect effectively
the combination of the usual plane-wave particle with the description through the spherical harmonics, the twisted particles have a defined propagation direction together with the
essential orbital angular momentum (OAM) projections on the same propagation axis.
It opens a window for the direct investigation of AS-functions owing to the corresponding unique $\phi$-dependence of
the differential cross sections \cite{ACK}.

%%%%%%%%%%%%%%
\section*{Acknowledgements}
%%%%%%%%%%%%%%
Our special thank goes to  L.~Szymanowski who provides us
very useful and illuminating discussions.
Also, I.V.A. thanks his colleagues  from
the Sun Yat-Sen University for a very warm hospitality.
I.V.A. also appreciates stimulating discussions with 
S.~Bolotsky, K.~Dybo,  A.~Lunis, A.~Markov and K.~Nikitina.
The work has been supported in part by PIFI 2024PVA0110
Program.
This work is supported by National Key R$\&$D Program of China No.2024YFE0109800 and 2024YFE0109802.

%%%%%%%%%%%%%%%%%%%%%%%%%%%

\end{document}